\documentclass[12pt,draftclsnofoot,journal,onecolumn]{IEEEtran}
\usepackage{algorithm}
\usepackage{color}
\usepackage{algpseudocode}
\usepackage{graphics}
\usepackage{epsfig}
\usepackage{cite}
\usepackage{threeparttable}
\usepackage{graphicx}
\usepackage{picinpar}
\usepackage{amssymb}
\usepackage[cmex10]{amsmath}
\usepackage{changebar}
\usepackage{subfigure}
\usepackage{stfloats}

\usepackage{bm}
\usepackage{setspace}

\input epsf
\begin{document}

\title{\huge{Terahertz Wireless Communications with Flexible Index Modulation Aided Pilot Design} }

\author{Tianqi Mao and Zhaocheng Wang, \IEEEmembership{Fellow,~IEEE}
\thanks{Manuscript received July 7, 2020; revised November 22, 2020 and January 27, 2021; accepted March 1, 2021. This work was supported in part by the National Key R\&D Program of China under Grant 2018YFB1801501 and in part by National Natural Science Foundation of China (Grant No. 61871253). \emph{(Corresponding author: Zhaocheng Wang.)}}
\thanks{T. Mao and Z. Wang are with Beijing National Research Center for Information Science and Technology, Department of Electronic Engineering, Tsinghua University, Beijing 100084, China,
and Z. Wang is also with Tsinghua Shenzhen International Graduate School, Shenzhen 518055, China (e-mail: maotq18@mails.tsinghua.edu.cn, zcwang@tsinghua.edu.cn).} %
\vspace{-3mm}} %

\maketitle
\begin{abstract}
Terahertz (THz) wireless communication is envisioned as a promising technology, which is capable of providing ultra-high-rate transmission up to Terabit per second. However, some hardware imperfections, which are generally neglected in the existing literature concerning lower data rates and traditional operating frequencies, cannot be overlooked in the THz systems. Hardware imperfections usually consist of phase noise, in-phase/quadrature imbalance, and nonlinearity of power amplifier. Due to the time-variant characteristic of phase noise, frequent pilot insertion is required, leading to decreased spectral efficiency. In this paper, to address this issue, a novel pilot design strategy is proposed based on index modulation (IM), where the positions of pilots are flexibly changed in the data frame, and additional information bits can be conveyed by indices of pilots. Furthermore, a turbo receiving algorithm is developed, which jointly performs the detection of pilot indices and channel estimation in an iterative manner. It is shown that the proposed turbo receiver works well even under the situation where the prior knowledge of channel state information is outdated. Analytical and simulation results validate that the proposed schemes achieve significant enhancement of bit-error rate performance and channel estimation accuracy, whilst attaining higher spectral efficiency in comparison with its classical counterpart.
\end{abstract}

\section{Introduction}
With the escalating requirement of wireless data throughput and proliferation of multimedia terminals, exploitation of Terahertz (THz) spectrum has attracted enormous attentions from the global community due to its potential of achieving terabit-per-second (Tbps) data transmission \cite{Alouini_survey_2020,Ian_PC_14,PC_survey_19,Chen_survey_19,Xianbinyu_conf_16,zcwang_mag_11}. This motivates the emergence of numerous bandwidth-consuming applications, such as high-definition holographic video meeting, virtual/augmented reality, and wireless backhaul, etc. \cite{Ian_PC_14,PC_survey_19,Boulogeorgos_mag_18}. Besides, THz beams are extremely narrow in order to overcome the severe signal attenuation, which adds difficulty to eavesdropping in the wiretap scenario, enabling secure data transmission for military applications \cite{Ian_PC_14}. Furthermore, THz waves are also considered for joint radar and communication (JRC) systems due to its desirable resolution and highly directional beams \cite{Alouini_survey_2020}. Thanks to the considerable merits above, THz wireless communication has been regarded as a promising candidate for the next-generation networks \cite{Rappaport_survey_19}.

However, there still exist tremendous challenges for commercialization of THz wireless communications. Firstly, in comparison with communication systems operating at lower-frequency bands, THz signals suffer from more severe attenuation induced by free-space path loss and molecular absorption effects from atmospheric particles like $\text{H}_2\text{O}$ \cite{Song_tts_11}. To address this issue, high-gain directional antennas (or antenna arrays with beamforming) are often invoked to maintain both the coverage and the quality of THz signals \cite{Rappaport_survey_19,Papasotiriou_cl_20}. Aside from the channel peculiarities, another critical point that limits the performance of THz wireless communications is hardware imperfection, which includes local-oscillator (LO) phase noise, in-phase/quadrature (I/Q) imbalance, and nonlinearity of the power amplifier \cite{Schenk_book_08,Boulogeorgos_access_19}. These impairments are generally omitted in low-rate systems by assuming ideal RF front ends \cite{Bjornson_tcom_13}. However, they cannot be overlooked in THz communication systems due to the well-known ``THz Gap'' posing considerable difficulty to THz hardware fabrication \cite{Alouini_survey_2020}, which have a detrimental impact on the performance of THz communication systems \cite{Boulogeorgos_access_19}\footnote{Granting that the hardware impairments can be mitigated with compensation techniques, there still exists residual imperfections that cannot be fully eliminated. Neglecting these residual imperfections could cause substantial performance loss \cite{Bjornson_tcom_13,Studer_conf_10}.}. Hence, researchers have recently started to investigate the hardware imperfection issue regarding THz communication systems. In \cite{Boulogeorgos_access_19}, distortions induced by the hardware imperfections were approximately modelled as circularly symmetric complex Gaussian (CSCG) variables. This model was further introduced to the beamforming-aided multiple-input single-output (MISO) and THz spatial modulation (THz-SM) systems in \cite{Boulogeorgos_wcnc_19} and \cite{Mao_tvt_20}, respectively, where the deleterious impact of hardware impairments is validated, and the corresponding receiving algorithms were developed to mitigate the resultant performance degradation. Except for imperfections of the mixer, the LO and the amplifier, additional nonlinearity induced by the frequency-multiplier-last architecture for THz wave generation was also considered in \cite{Ramadan_access_18} and \cite{Mao_icc_20}, where pre-compensation schemes and modified receiver design were proposed correspondingly.

\begin{figure*}[t!]
\begin{center}
\includegraphics[width=1\linewidth, keepaspectratio]{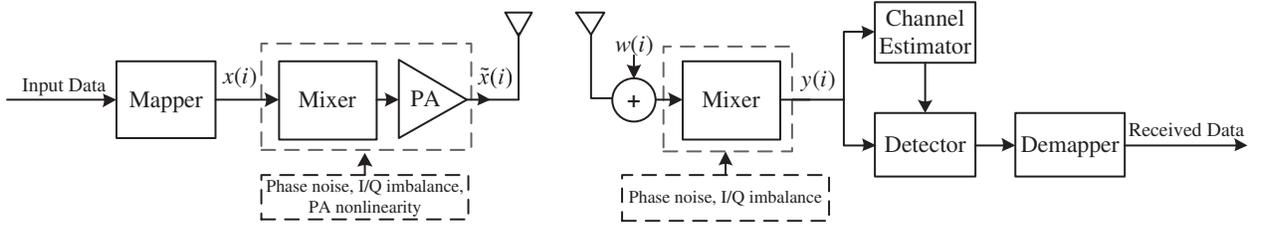}
\end{center}
\caption{Transceiver diagram of the THz single-carrier system with hardware imperfections.}
\label{fig1}
\end{figure*}
Aside from the performance loss caused by hardware imperfections, due to the time-variant property of strong phase noise mentioned in \cite{Colavolpe_jsac_05} and \cite{Krishnan_tcom_13}, channel-tracking pilots are required to be frequently utilized in order to timely update the equivalent channel fading coefficients in THz communication systems. Conventional channel estimation methods usually insert preamble sequences with fixed positions periodically in the data frame \cite{Jiang_cl_13,Buzzi_tvt_04}, which causes undesirable spectral efficiency loss\footnote{Although large bandwidth is usually employed in THz wireless communications, the system spectral efficiency has to be sufficiently high in order to achieve over 100 Gbps data rate. Besides, only slight spectral efficiency loss could result in considerable data rate degradation in THz communication systems with tens of GHz bandwidth.}. In this paper, a THz-band single-carrier communication system under strong hardware impairments is investigated. To compensate for the throughput loss caused by frequent channel tracking, whilst attaining high channel estimation accuracy simultaneously, index modulation (IM) is introduced to pilot design, which is capable of achieving superior spectrum and energy efficiency gains for wireless communication systems \cite{Basar_survey_17,Mao_survey_19,Basar_tsp_13,Wen_mm_17,Li_globalsip_18}. {More specifically, aside from the data symbol bits, additional information can be implicitly conveyed by the subcarrier activation patterns in orthogonal frequency division multiplexing (OFDM) systems, e.g., the ``on/off'' status of subcarriers \cite{Basar_tsp_13}, index of used constellation alphabet for each subcarrier \cite{Wen_mm_17}, and positions of inserted pilots in the frequency domain \cite{Li_globalsip_18}.} Based on the IM philosophy, a novel flexible pilot design strategy is proposed. More specifically, unlike classical training sequences with fixed position, the IM-aided pilots are flexibly inserted into the data frame, and their positions are determined by the conveyed information bits, referred to as \emph{index bits}. In other words, additional binary bits can be conveyed by the indices of pilots, which cancels out the throughput loss, and even achieves higher spectral efficiency than data transmission without pilot sequences. To detect the indices of pilot symbols at the receiver, referred to as \emph{index pattern}, distinguishable constellation alphabets are employed for the pilots and data symbols, respectively. Then a turbo receiving algorithm is proposed by updating the index pattern of pilots and the estimate of channel state information (CSI) iteratively, which ensures superior precision of index pattern detection and channel estimation. {Furthermore, the extension of the proposed flexible IM-aided pilot design to frequency-selective fading channels considering THz communication systems with ultra-broad bandwidth, is briefly discussed.} To validate the feasibility of the proposed turbo receiving algorithm, intuitive performance analysis is carried out from a geometrical point of view, followed by complexity evaluation. Simulation results demonstrate that the proposed flexible IM-aided pilot design and the corresponding turbo receiving algorithm are capable of enhancing both the channel estimation accuracy and bit-error rate (BER) performance at higher spectral efficiency, compared with its conventional counterpart with fixed pilot assignment. Analytical and simulative results also show that the proposed turbo receiving algorithm still works well when the prior knowledge of CSI fed into the iterative receiver is outdated.

The remainder of the paper is organized as follows. Section \ref{s2} presents the transceiver model of the THz single-carrier communication system with hardware imperfections. Then the proposed flexible IM-aided pilots and the turbo receiving algorithm are illustrated in Section \ref{s3}. Afterwards, performance analysis and simulation results are provided in Section \ref{s4} and \ref{s5}, respectively. Finally, Section \ref{s6} draws the conclusion.

\emph{Notation:} $[\cdot]^T$ and $[\cdot]^H$ denote the transpose and the conjugate transpose. $\left \|\cdot \right \|$, $\left \lfloor \cdot \right \rfloor$ and $|\cdot|$ stand for the 2-norm, the floor operator, and the length of a line segment, respectively. Besides, $\binom{\cdot}{\cdot}$ represents the binomial coefficient.
\begin{figure*}[t!]
\centering
\subfigure[Frame structure with classical fixed preamble.]{
\begin{minipage}[b]{0.65\textwidth}
\includegraphics[width=1\textwidth]{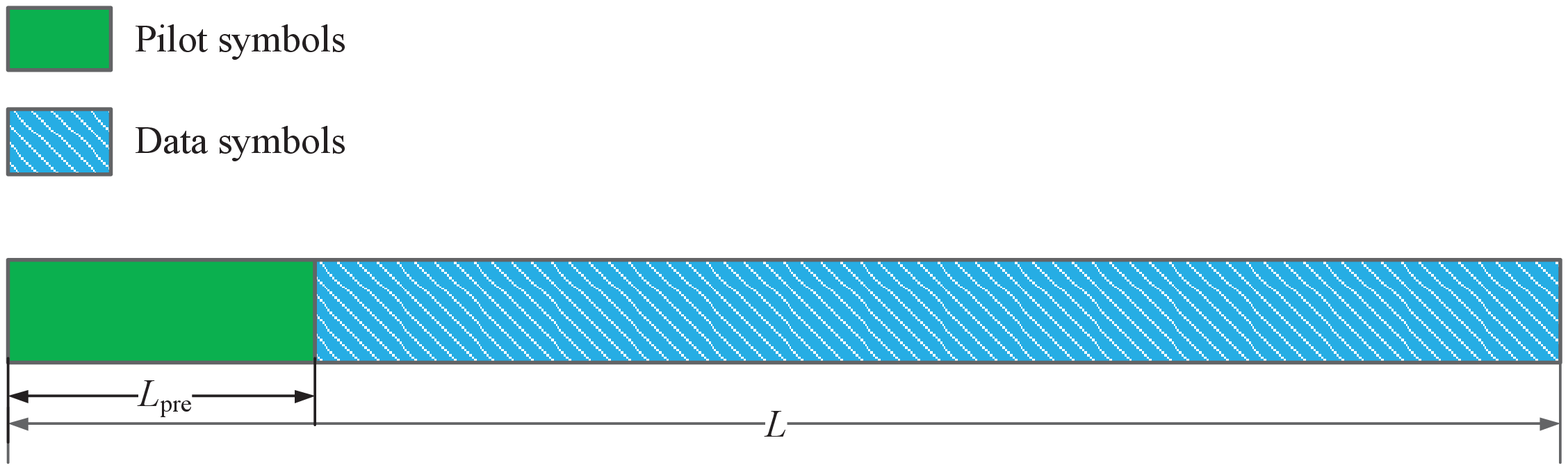} \\
\end{minipage}
}
\subfigure[Frame structure with the proposed flexible IM-aided pilots.]{
\begin{minipage}[b]{0.65\textwidth}
\includegraphics[width=1\textwidth]{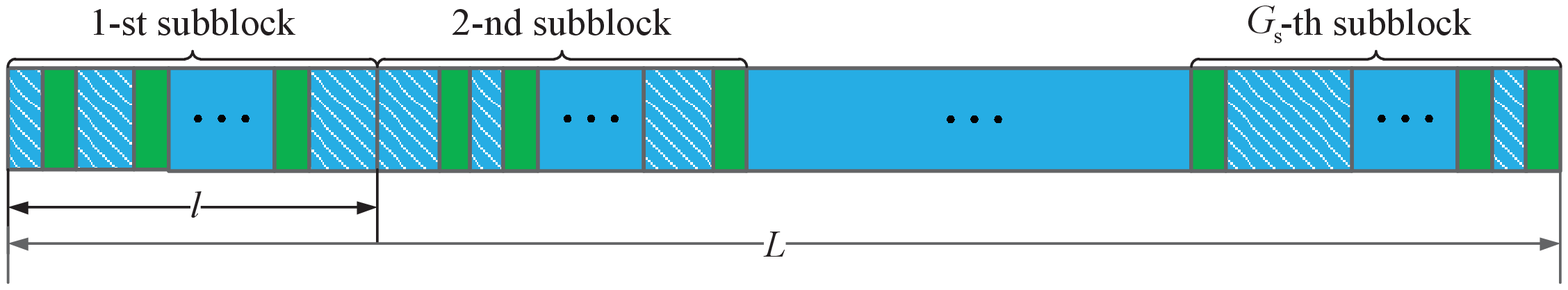} \\
\end{minipage}
}
\caption{Frame structure with classical fixed preamble and the proposed flexible IM-aided pilots.}
\label{fig2}
\end{figure*}
\section{System Model}\label{s2}
The transceiver model of the THz single-carrier communication system is illustrated in Fig. \ref{fig1}. {Note that aside from the use of directional antennas, large antenna arrays with beamforming could be adopted to compensate for the severe path loss of THz signals, which are actually equivalent to a single high-gain directional antenna if the same signal is transmitted on each antenna element for diversity gain. Hence, a single-input single-output (SISO) scenario is assumed in this paper for simplicity.} At the transmitter, the incoming bits are firstly modulated as $M$-ary data symbols, which are then combined with the pilots, generating an $N$-dimensional signal vector denoted as $\mathbf{x}=[x(1),x(2),\cdots,x(N)]^T$ with average transmit power $P_\text{t}$. Afterwards, $x(i)$ for $i=1,2,\cdots,N$ is processed by the mixer and power amplifier to generate THz waves $\tilde{x}(i)$ for data transmission, which induces non-negligible phase noise, I/Q imbalance and nonlinearity, leading to severe signal distortions at the transmitter side. For simplicity, the nonlinearity of the power amplifier is assumed to be already compensated \cite{Qi_tvt_10,Shabany_tcs_08}. According to the mathematical models for phase noise and I/Q imbalance \cite{Colavolpe_jsac_05,Jiang_cl_13}, the distorted signal $\tilde{x}(i)$ can be formulated as
\begin{equation}\label{eq1}
\tilde{x}(i)=\left ( \mu_\text{t}x(i)+v_\text{t}x^*(i) \right )e^{j\theta_\text{t}(i)},\:\:i=1,2,\cdots,N,
\end{equation}
where we have
\begin{equation}\label{eq2}
\left\{\begin{matrix}
\mu_\text{t}=\cos\phi_\text{t}-j\epsilon_\text{t}\sin\phi_\text{t}\\
v_\text{t}=\epsilon_\text{t}\cos\phi_\text{t}-j\sin\phi_\text{t}
\end{matrix}.\right.
\end{equation}
$\epsilon_\text{t}$ and $\phi_\text{t}$ denote the amplitude and phase imbalance of the I/Q branches, respectively. Besides, $\theta_\text{t}(i)$ is the phase noise component at the transmitter. {For ultra-high-rate THz communication systems, the impacts of phase noise on transmitted signals are considered to be frequency-dependent \cite{Schenk_book_08,Steendam_phd_00}. On one hand, the lower-frequency components in the power spectral density (PSD) of phase noise contribute to phase errors of the THz signals; On the other hand, the higher-frequency components cause inter-carrier interference (ICI) to the adjacent carriers \cite{Papasotiriou_cl_20}. Since we focus on a single-carrier THz system in this paper, only the rotational effects of phase noise are taken into consideration as illustrated in (\ref{eq2}).} In \cite{Colavolpe_jsac_05}, the process of $\theta_\text{t}(i)$ for $i=1,2,\cdots,N$ follows a random-walk model, where the phase noise changes in a symbol-wise manner. However, for THz wireless communications, since ultra-wide bandwidth is usually employed, leading to extremely short symbol period, it is reasonable to assume a block-wise changing model for the phase noise, which keeps constant within each data block, whilst varies over adjacent blocks randomly. More specifically, the data frame $\mathbf{x}$ is equally separated into $L$-length blocks $\mathbf{x}_{k}=\left[x({(k-1)L+1}),x({(k-1)L+2}),\cdots,x({kL})\right]^T$ for $k=1,2,\cdots,G$, where $G$ equals $N/L$. Then phase noise terms within each data block are assumed to be constant, shown as
\begin{equation}\label{eq3}
\theta_\text{t}((k-1)L+1)=\theta_\text{t}((k-1)L+2)=\cdots=\theta_\text{t}(kL)\overset{\underset{\mathrm{def}}{}}{=}\theta_{\text{t},k},
\end{equation}
whilst the phase noise for adjacent data blocks follow the random-walk model, expressed by
\begin{equation}\label{eq4}
\theta_{\text{t},k+1}=\theta_{\text{t},k}+\Delta\theta_{\text{t},k},\:\:k=1,2,\cdots,G-1,
\end{equation}
where $\theta_{\text{t},1}$ is uniformly distributed in $[0,2\pi)$, and $\Delta\theta_{t,k}$ represents the random variation of phase noise between the $k$-th and $(k+1)$-th data blocks, following Gaussian distribution as $\mathcal{N}(0,\sigma_\Delta^2)$.

To mitigate the severe signal attenuation induced by path loss and molecular absorption effects, high-gain directional antennas are employed at the THz transceiver, which indicates that multi-path effects are negligible. Therefore, a line-of-sight (LoS) flat-fading channel is assumed \cite{Han_tsp_16}\footnote{It has been stated in \cite{Han_tsp_16} that the coherence bandwidth exceeds 60 GHz for THz communication systems at 300 GHz equipped with directional antennas. Therefore, the frequency-selective fading due to spreading and molecular absorption loss is not mainly considered in this paper.}, denoted as $h(i)$ for $i=1,2,\cdots,N$. {Based on the THz channel characteristics \cite{Boulogeorgos_access_19}, the overall path gain $\|h(i)\|$ can be expressed as
\begin{equation}
\|h(i)\|=h_{s}(i)h_{a}(i)h_{m}(i),\:\:i=1,2,\cdots,N,
\end{equation}
where $h_s(i)$ and $h_a(i)$ denote the spreading loss and molecular absorption loss, and $h_m(i)$ is the beam misalignment fading originated from the narrow beam-width of directional antennas. The expressions of these coefficients are detailed in \cite{Boulogeorgos_access_19}, thus omitted here for brevity.}

According to Fig. \ref{fig1}, aside from the additive white Gaussian noise (AWGN) component $w(i)$ following $\mathcal{CN}(0,\sigma^2)$, additional phase noise and I/Q imbalance are induced at the receiver side. For simplicity, these imperfections are well approximated as an additive distortion term $n_\text{r}(i)$ following $\mathcal{CN}(0,\kappa^2P_\text{r})$, where $P_\text{r}$ is the average power of the received signals, and $\kappa^2$ denotes the level of hardware imperfections at the receiver \cite{Bjornson_tcom_13,Boulogeorgos_access_19}. Then the distorted signal $y(i)$ for $i=1,2\cdots,N$ can be formulated as\footnote{In this paper, the approximate distortion model is only applied to characterize the hardware imperfections at the receiver, because it is consider to be less dominant than the counterpart at the transmitter, as explained in \cite{Zhang_tcom_15}.}
\begin{equation}\label{eq5}
\begin{aligned}
y(i)=h(i)\tilde{x}(i)+n_\text{r}(i)+w(i).
\end{aligned}
\end{equation}
By substituting (\ref{eq1}) into (\ref{eq5}), the expression of $y(i)$ can be further derived as
\begin{equation}\label{eq8}
y(i)=\left [ x(i),x(i)^* \right ]\begin{bmatrix}
h(i)\mu_{\text{t}}e^{j\theta_{\text{t}}(i)}\\
h(i)v_{\text{t}}e^{j\theta_{\text{t}}(i)}
\end{bmatrix}+\tilde{w}(i),
\end{equation}
where $\begin{bmatrix}
h(i)\mu_{\text{t}}e^{j\theta_{\text{t}}(i)}\\
h(i)v_{\text{t}}e^{j\theta_{\text{t}}(i)}
\end{bmatrix}$ is the equivalent channel fading vector, and $\tilde{w}(i)=n_\text{r}(i)+w(i)$ denotes the distortion-plus-noise component, following $\mathcal{CN}(0,\kappa^2P_\text{r}+\sigma^2)$. Finally, $y(i)$ is processed by the signal detector and a constellation demapper to demodulate the information bits, where pilot-assisted channel estimation is performed.
\begin{figure*}[t!]
\begin{center}
\includegraphics[width=0.65\linewidth, keepaspectratio]{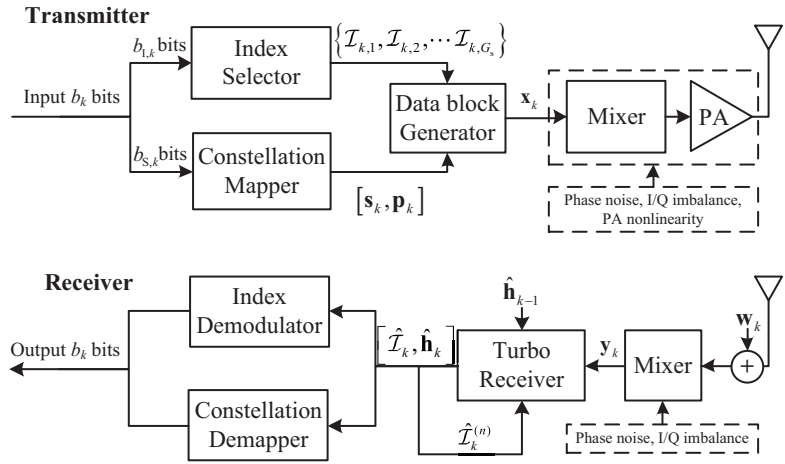}
\end{center}
\caption{Transceiver architecture of the THz single-carrier system using the proposed IM-aided flexible pilots and turbo receiving algorithm.}
\label{fig3}
\end{figure*}
\section{Proposed Schemes}\label{s3}
Based on the block-wise random-walk property of phase noise, it is assumed that the equivalent channel fading vector in (\ref{eq8}) remains approximately constant within each data block, denoted as $\mathbf{h}_k$ for $k=1,2,\cdots,G$, whilst changes happen across different blocks due to random fluctuations of phase noise as well as possible variations of the physical channel coefficients.
\subsection{Classical Pilot Assignment}
Classical channel estimation methods usually employ a preamble sequence inserted at the beginning of each data block \cite{Jiang_cl_13,Buzzi_tvt_04}, as illustrated in Fig. \ref{fig2} (a), where $L_{\text{pre}}$ denotes the preamble length for each data block. Then the channel estimate of the $k$-th data block can be obtained with least-squared (LS) algorithm as \cite{Kay_book_93}
\begin{equation}\label{LS}
\hat{\mathbf{h}}_{\text{conv},k}=(\mathbf{P}_{k,\text{pre}}^H\mathbf{P}_{k,\text{pre}})^{-1}\mathbf{P}_{k,\text{pre}}^H\mathbf{y}_{\text{p},k},\:\:k=1,2,\cdots,G,
\end{equation}
where $\mathbf{P}_{k}$ is the $L_{\text{pre}}\times 2$ preamble matrix consisting of the pilot symbols and their conjugates, formulated as
\begin{equation}
\mathbf{P}_{k,\text{pre}}=\begin{bmatrix}
p_{k}(1),p_{k}(2),\cdots,p_{k}(L_{\text{pre}})\\
p_{k}^*(1),p_{k}^*(2),\cdots,p_{k}^*(L_{\text{pre}})
\end{bmatrix}^T,
\end{equation}
and $\mathbf{y}_{\text{p},k}$ is defined as the corresponding received signal vector. Based on the block structure in Fig. \ref{fig2} (a), the system spectral efficiency can be calculated as
\begin{equation}\label{eq_se_conv}
\text{SE}_{\text{conv}}=\frac{L-L_{\text{pre}}}{L}\log_2(M).
\end{equation}
It is readily seen from (\ref{eq_se_conv}) that the classical fixed preamble results in throughput degradation.
\subsection{Proposed Flexible IM-Aided Pilot Design}\label{s3b}
\begin{table}[t!]
\vspace*{-1mm}
\renewcommand{\arraystretch}{1.1}
\caption{A Look-up Table Between the Index Bits and Indices of the Flexible IM-Aided Pilots for $l=4$ and $l_\text{p}=2$.}
\begin{center}
\begin{tabular}{| c| c |c|}
\hline
 Index bits & Indices & Subblocks \\\hline
 $[0,0]$    & $\{1,2\}$ & $\big[p_{k,g}(1),p_{k,g}(2),s_{k,g}(1),s_{k,g}(2)\big]$ \\\hline
 $[0,1]$    & $\{2,3\}$ & $\big[s_{k,g}(1),p_{k,g}(1),p_{k,g}(2),s_{k,g}(2)\big]$ \\\hline
 $[1,0]$    & $\{3,4\}$ & $\big[s_{k,g}(1),s_{k,g}(2),p_{k,g}(1),p_{k,g}(2)\big]$ \\\hline
 $[1,1]$    & $\{1,4\}$ & $\big[p_{k,g}(1),s_{k,g}(1),s_{k,g}(2),p_{k,g}(2)\big]$\\\hline
\end{tabular}
\end{center}
\vspace*{-3mm}
\label{t1}
\end{table}
To address this issue, a flexible pilot design strategy based on IM is proposed for the THz communication system, as illustrated in Fig. \ref{fig2} (b) and Fig. \ref{fig3}. At the transmitter, the incoming $b_k$ bits for the $k$-th data block is firstly divided into $b_{\text{S},k}$ data symbol bits and $b_{\text{I},k}$ index bits. On one hand, the $b_{\text{S},k}$ binary bits are fed into a constellation mapper of an $M_{\text{s}}$-ary alphabet $\mathcal{M}_{\text{s}}=\{S(1),S(2),\cdots,S(M_{\text{s}})\}$, yielding the data symbol vector $\mathbf{s}_{k}=\left[s_{k}(1),s_{k}(2),\cdots,s_{k}(L_{\text{s}})\right]^T$. On the other hand, $L_{\text{p}}$ pilots $\mathbf{p}_{k}=\left[p_{k}(1),p_{k}(2),\cdots,p_{k}(L_{\text{p}})\right]^T$ are generated using an $M_{\text{p}}$-ary constellation alphabet $\mathcal{M}_{\text{p}}=\{S_{\text{p}}(1),S_{\text{p}}(2),\cdots,S_{\text{p}}(M_{\text{p}})\}$ for channel estimation in each data block. Note that $L=L_{\text{s}}+L_{\text{p}}$. Unlike classical preamble assignment, the proposed IM-aided pilots are no longer fixed, whose indices in the data block are determined by the $b_{\text{I},k}$ index bits with an index selector. More specifically, as presented in Fig. \ref{fig2} (b), the $k$-th data block $\mathbf{x}_{k}$ for $k=1,2,\cdots,G$ is further split into $G_{\text{s}}$ subblocks with size of $l=L/G_{\text{s}}$, denoted as $\mathbf{x}_{k,g}=\left[x_{k,g}(1),x_{k,g}(2),\cdots,x_{k,g}(l)\right ]^T$ for $g=1,2,\cdots,G_{\text{s}}$. In each subblock, aside from the $l_{\text{s}}$ data symbols, there are totally $l_\text{p}$ pilots drawn from $\mathbf{p}_k$, denoted as $\mathbf{p}_{k,g}=\left[p_{k,g}(1),p_{k,g}(2),\cdots,p_{k,g}(l_{\text{p}})\right]^T$, corresponding to the index set $\mathcal{I}_{k,g}=\{i_{k,g}(1),i_{k,g}(2),\cdots,i_{k,g}(l_\text{p})\}$. Here $\mathcal{I}_{k,g}$ is determined by the incoming bits according to the mapping table between the index bits and the pilot positions, satisfying $1\leq i_{k,g}(1)<i_{k,g}(2)<\cdots<i_{k,g}(l_\text{p})\leq l$, which can be exemplified by Table \ref{t1} with $l=4$ and $l_\text{p}=2$.
%
By such arrangement, additional $\left \lfloor \log_2\binom{l}{l_\text{p}} \right \rfloor$ bits are conveyed by the pilot indices in each subblock. Hence, the system spectral efficiency applying the proposed flexible IM-aided pilots becomes
\begin{equation}
\text{SE}_{\text{prop}}=\frac{1}{l}\left ((l-l_\text{p})\log_2(M_{\text{s}})+\left \lfloor \log_2\binom{l}{l_\text{p}} \right \rfloor \right ),
\end{equation}
which indicates that the proposed pilot design is capable of compensating for the throughput loss induced by pilots, and even attaining higher spectral efficiency than transmission without using pilots, given that $\left \lfloor \log_2\binom{l}{l_\text{p}} \right \rfloor>l_{\text{p}}\log_2(M_\text{s})$.
\begin{figure}[t!]
\begin{center}
\includegraphics[width=0.8\linewidth, keepaspectratio]{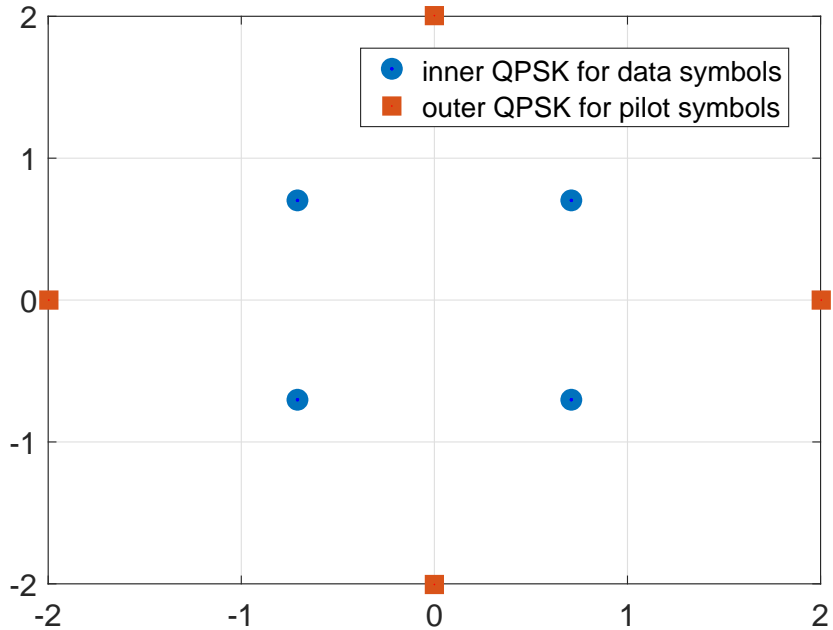}
\end{center}
\caption{Different constellation alphabets employed for pilots and data symbols.}
\label{fig4}
\vspace*{-3mm}
\end{figure}

{In order to detect the indices of pilots for channel estimation and index bit recovery, the constellation alphabets employed for data symbols and pilots should be distinguishable, i.e., $\mathcal{M}_{\text{s}}\cap  \mathcal{M}_{\text{p}}=\varnothing$. As an example, by considering the phase-shift-keying (PSK) format, one feasible constellation design for $M_{\text{s}}=M_{\text{p}}=4$ is illustrated in Fig. \ref{fig4}, where $\mathcal{M}_\text{s}=\left \{e^{j\frac{1}{4}\pi},e^{j\frac{3}{4}\pi},e^{j\frac{5}{4}\pi},e^{j\frac{7}{4}\pi}\right \}$ and $\mathcal{M}_\text{p}=\left \{\sqrt{\gamma},\sqrt{\gamma}j,-\sqrt{\gamma},-\sqrt{\gamma}j \right \}$. Note that $\gamma$ denotes the average power ratio between $\mathcal{M}_{\text{p}}$ and $\mathcal{M}_{\text{s}}$, and is set as $4$ in this example, which however is not guaranteed to be optimal. The impacts of $\gamma$ on the system BER performance are discussed as follows.

Firstly, the normalized minimum Euclidean distance between constellations in $\mathcal{M}_{\text{s}}$ and $\mathcal{M}_{\text{p}}$, which determines the performance of index pattern detection \cite{Mao_survey_19}, can be expressed as
\begin{equation}\label{dmin_gamma_cal}
d_{\min}=2-\frac{4}{\frac{2}{\sqrt{\gamma}}+\sqrt{\gamma}},
\end{equation}
whose minimum value equals $2-\sqrt{2}$ as $\gamma=2$. Then it can be seen that, the value of $\gamma$ should be sufficiently away from $2$ to enlarge $d_{\min}$, thus improving the accuracy of the index pattern detection. Besides, with the increase of $\gamma$, the average transmit power of pilots becomes larger, which enhances the channel estimation accuracy. Note that for our proposed flexible IM-aided pilot design, channel estimation errors not only negatively affect demodulation of data symbols, but also contribute to poor index pattern detection, which leads to increased BER of index bits, and further causes additional errors of data symbol bits due to wrong detections of the index pattern. Hence, $\gamma$ needs to be sufficiently large for accurate channel estimation.

Furthermore, in the $k$-th data block, by assuming that its channel estimation error is tolerable to correctly detect the index pattern, the channel estimate using LS criterion can be expressed as
\begin{equation}
\hat{\mathbf{h}}_k=\mathbf{h}_k+(\mathbf{P}^H\mathbf{P})^{-1}\mathbf{P}^H\tilde{\mathbf{w}}_k,
\end{equation}
where $\mathbf{P}_k=\left [\mathbf{p}_k,\text{conj}(\mathbf{p}_k) \right ]$, and $\tilde{\mathbf{w}}_k$ denotes the distortion-plus-noise term following $\mathcal(0,(\kappa^2P_\text{r}+\sigma^2)\mathbf{I}_{L_\text{p}})$. Here $\mathbf{I}_{L_\text{p}}$ is the $L_\text{p}$-by-$L_\text{p}$ identity matrix. By assuming that each candidate of $\mathcal{M}_{\text{p}}$ is chosen as the pilot symbol equiprobably, the channel estimation error, defined as $\Delta\mathbf{h}=(\mathbf{P}^H\mathbf{P})^{-1}\mathbf{P}^H\tilde{\mathbf{w}}_k$, approximately follows CSCG distribution as $\mathcal{CN}\left(0,\frac{L_\text{p}\gamma+L_\text{s}}{LL_\text{p}\gamma}(\kappa^2P_\text{r}+\sigma^2)\mathbf{I}_2\right)$, where the transmitted power is set as $1$ without loss of generality. Afterwards, the received signal for data symbol detection can be formulated as
\begin{equation}
\begin{aligned}
y_k(i)&=\sqrt{\frac{L}{L_\text{p}\gamma+L_{\text{s}}}}\left [ s_k(i),s_k^*(i) \right ](\hat{\mathbf{h}}_k-\Delta\mathbf{h})+\tilde{w}_k(i)\\&=\sqrt{\frac{L}{L_\text{p}\gamma+L_{\text{s}}}}\left [ s_k(i),s_k^*(i) \right ]\hat{\mathbf{h}}_k+{\tilde{w}}'_k(i),
\end{aligned}
\end{equation}
where we have the equivalent noise term written as
\begin{equation}
{\tilde{w}}'_k(i)=-\sqrt{\frac{L}{L_\text{p}\gamma+L_{\text{s}}}}\left [ s_k(i),s_k^*(i) \right ]\Delta\mathbf{h}+\tilde{w}_k(i),
\end{equation}
following $\mathcal{CN}\left(0, (\frac{2}{L_\text{p}\gamma}+1)(\kappa^2P_\text{r}+\sigma^2)\right)$. Then the received signal-to-noise ratio (SNR) can be approximately written as
\begin{equation}\label{snr_approx}
\text{SNR}\approx \frac{LP_\text{r}}{\left (L_\text{p}\gamma+2  \right )\left ( 1+\frac{L_\text{s}}{L_\text{p}\gamma} \right )(\kappa^2P_\text{r}+\sigma^2)}.
\end{equation}
It can be seen in (\ref{snr_approx}) that, as $\gamma$ increases to infinity, the SNR value approaches zero, causing significant BER performance degradation. Hence, an intuitive guideline for the power allocation issue is that, $\gamma$ should be set to be sufficiently over $2$ to ensure accuracy of channel estimation and index pattern detection, but cannot be too large in case of undesirable SNR degradation. This finding can be well supported by Fig. \ref{fig10}, where the system BER performance versus $\gamma$ is simulated. Note that a sophisticated trade-off is required in this optimal constellation design (power allocation) of $\mathcal{M}_{\text{p}}$ and $\mathcal{M}_{\text{s}}$ for the PSK format as Fig. \ref{fig4}, which may be out of the scope of the main idea in this paper. Hence, the derivative work of the optimal power allocation issue is set aside as our future work for detailed investigation. Instead, numerical methods can be applied to obtain the value of $\gamma$ that optimizes the BER performance as illustrated in Fig. \ref{fig10}.}
\subsection{Turbo Receiving Algorithm for Index Pattern Detection and Channel Estimation}\label{s3.3}
As shown in Fig. \ref{fig3}, a turbo receiving algorithm is developed to jointly estimate the index pattern and the channel fading vector in an iterative manner, which can be clearly illustrated with mathematical induction as below.

{\bf Initialization:}
Before data transmission, a preamble $\mathbf{p}_0=\left[p_0(1),p_0(2),\cdots,p_0({L}'_{\text{pre}})\right]^T$ of length ${L}'_{\text{pre}}$, is utilized to obtain an initial channel estimate $\hat{\mathbf{h}}_0$ with LS algorithm. Then $\hat{\mathbf{h}}_0$ is employed as the prior knowledge of CSI to detect the index pattern of pilots for the 1-st data block.

{\bf Coarse index pattern detection:}
For the $k$-th data block ($1\leq k\leq N$), $\hat{\mathbf{h}}_{k-1}$ is employed as its prior knowledge of CSI. The received signals in the $g$-th data subblock can be formulated as
\begin{equation}\label{subblock_rx}
y_{k,g}(i)=\left [ x_{k,g}(i),x_{k,g}(i)^* \right ]\mathbf{h}_k+\tilde{w}_{k,g}(i),\:\:i=1,2,\cdots,l,
\end{equation}
where
\begin{equation}
\begin{aligned}
&y_{k,g}(i)=y\left({(k-1)L+(g-1)l+i}\right),\\
&x_{k,g}(i)=x\left({(k-1)L+(g-1)l+i}\right),\\
&\tilde{w}_{k,g}(i)=\tilde{w}\left({(k-1)L+(g-1)l+i}\right).
\end{aligned}
\end{equation}
Here $\tilde{w}_{k,g}(i)$ follows CSCG distribution as $\mathcal{CN}(0,\kappa^2P_\text{r}+\sigma^2)$. Based on (\ref{subblock_rx}), a symbol-wise log-likelihood ratio (LLR) detector can be applied to obtain the positions of pilots \cite{Mao_survey_19}, where the logarithm ratio between \emph{a posteriori} probabilities of the event that the symbol is modulated by $\mathcal{M}_{\text{p}}$ or $\mathcal{M}_{\text{s}}$ is calculated as
\begin{equation}\label{llr1}
\eta_{k,g}(i)=\log\left ( \frac{\sum_{m=1}^{M_{\text{p}}}\Pr\left(x_{k,g}(i)=S_{\text{p}}(m)|y_{k,g}(i)\right)}{\sum_{n=1}^{M_{\text{s}}}\Pr\left(x_{k,g}(i)=S(n)|y_{k,g}(i)\right) } \right ).
\end{equation}
Since we have $\Pr\left ( x_{k,g}(i)=S_{\text{p}}(m) \right )=\frac{l_\text{p}}{lM_{\text{p}}}$ and $\Pr\left ( x_{k,g}(i)=S(n) \right )=\frac{l-l_\text{p}}{lM_{\text{s}}}$, (\ref{llr1}) can be further derived as
\begin{equation}\label{llr2}
\begin{aligned}
\eta_{k,g}(i)=\log\left ( \frac{l_\text{p}M_{\text{s}}}{M_{\text{p}}(l-l_\text{p})} \right )+\log\left ( \frac{\sum_{m=1}^{M_{\text{p}}}\Pr\left(y_{k,g}(i)|x_{k,g}(i)=S_{\text{p}}(m)\right)}{\sum_{n=1}^{M_{\text{s}}}\Pr\left(y_{k,g}(i)|x_{k,g}(i)=S(n)\right) } \right ),
\end{aligned}
\end{equation}
where the likelihood function $\Pr\left(y_{k,g}(i)|x_{k,g}(i)\right)$ is formulated as
\begin{equation}\label{likelihood}
\begin{aligned}
\Pr\left(y_{k,g}(i)|x_{k,g}(i)\right)=\frac{1}{\pi(\kappa^2{P}_\text{r}+\sigma^2)}\times\exp\left ( -\frac{\left \|y_{k,g}(i)- \left [ x_{k,g}(i),x_{k,g}(i)^* \right ]\mathbf{h}_{k} \right \|^2}{\kappa^2{P}_\text{r}+\sigma^2} \right ).
\end{aligned}
\end{equation}
By substituting (\ref{likelihood}) into (\ref{llr2}), the final expression of the LLR value can be derived as
\begin{equation}\label{llr3}
\begin{aligned}
&\eta_{k,g}(i,\hat{\mathbf{h}}_{k-1})=\log\left ( \frac{l_\text{p}M_{\text{s}}}{M_\text{p}(l-l_\text{p})}\right)+\log\left (\sum_{m=1}^{M_\text{p}}\exp\left ( -\frac{\left \|y_{k,g}(i)- \left [ S_\text{p}(m),S^*_\text{p}(m) \right ]\hat{\mathbf{h}}_{k-1} \right \|^2}{\kappa^2\hat{P}_\text{r}+\sigma^2} \right )  \right ) -\\&\log\left (\sum_{n=1}^{M_\text{s}}\exp\left ( -\frac{\left \|y_{k,g}(i)- \left [ S(n),S^*(n) \right ]\hat{\mathbf{h}}_{k-1} \right \|^2}{\kappa^2\hat{P}_\text{r}+\sigma^2} \right )  \right ),
\end{aligned}
\end{equation}
where $\eta_{k,g}(i,\hat{\mathbf{h}}_{k-1})$ stands for the LLR value calculated using $\hat{\mathbf{h}}_{k-1}$ as the channel estimate, and $\hat{P}_\text{r}=\hat{\mathbf{h}}_{k-1}^H\hat{\mathbf{h}}_{k-1}P_\text{t}$ is the estimate of $P_\text{r}$. After all the LLR values of the $g$-th subblock are obtained using (\ref{llr3}), the indices of pilots in each subblock can be coarsely determined by
\begin{equation}\label{index_cal}
\begin{aligned}
\hat{\mathcal{I}}_{k,g}^{(0)}&=\{\hat{i}_{k,g}^{(0)}(1),\hat{i}_{k,g}^{(0)}(2),\cdots,\hat{i}_{k,g}^{(0)}(l_\text{p})\}\\&=\arg \underset{{1\leq i\leq l}} {l_\text{p}\text{-}\max}\:\eta_{k,g}(i,\hat{\mathbf{h}}_{k-1}),\:\: g=1,2,\cdots,G_{\text{s}},
\end{aligned}
\end{equation}
where $\hat{i}_{k,g}^{(0)}(1)<\hat{i}_{k,g}^{(0)}(2)<\cdots<\hat{i}_{k,g}^{(0)}(l_\text{p})$, and the operator $l_\text{p}\text{-}\max$ finds the largest $l_\text{p}$ LLR values. Then the resultant $\hat{\mathcal{I}}^{(0)}_{k}=\{\hat{\mathcal{I}}_{k,1}^{(0)},\hat{\mathcal{I}}_{k,2}^{(0)},\cdots,\hat{\mathcal{I}}_{k,G_\text{s}}^{(0)}\}$ is defined as the coarse estimate of the pilot index pattern for the $k$-th data block.

{\bf Iterative joint index pattern detection and channel estimation:} At the $n$-th iteration ($n\geq 1$), the positions of the flexible IM-aided pilots are firstly determined by $\hat{\mathcal{I}}^{(n-1)}_{k}$, and then they are employed to update the channel estimate, where the resultant CSI is further fed back for updating the pilot index pattern. More specifically, for the $g$-th subblock, the received symbols of the indices $\hat{\mathcal{I}}^{(n-1)}_{k}\setminus \hat{\mathcal{I}}^{(n-1)}_{k,g}$ are concatenated together, as illustrated in (\ref{eq22}) (on the top of next page),
\begin{figure*}[t!]
\begin{align}\label{eq22}
\tilde{\mathbf{y}}_{k,g}=&[y_{k,1}(\hat{i}^{(n-1)}_{k,1}(1)),\cdots,y_{k,1}(\hat{i}^{(n-1)}_{k,1}(l_{\text{p}})),y_{k,2}(\hat{i}^{(n-1)}_{k,2}(1)),\cdots,
y_{k,g-1}(\hat{i}^{(n-1)}_{k,g-1}(1)),\cdots,y_{k,g-1}(\hat{i}^{(n-1)}_{k,g-1}(l_{\text{p}})),\nonumber\\&y_{k,g+1}(\hat{i}^{(n-1)}_{k,g+1}(1)),\cdots,y_{k,g+1}(\hat{i}^{(n-1)}_{k,g+1}(l_\text{p})),\cdots,y_{k,G_\text{s}}(\hat{i}^{(n-1)}_{k,G_\text{s}}(l_\text{p}))]^T.
\end{align}
\hrulefill
\end{figure*}
which is then utilized to estimate the channel fading vector
\begin{equation}\label{eq_23_ls}
\hat{\mathbf{h}}^{(n)}_{k,g}=(\tilde{\mathbf{P}}_{k,g}^H\tilde{\mathbf{P}}_{k,g})^{-1}\tilde{\mathbf{P}}_{k,g}^H\tilde{\mathbf{y}}_{k,g},
\end{equation}
where $\tilde{\mathbf{P}}_{k,g}$ is an $(L_\text{p}-l_\text{p})\times 2$ pilot matrix defined as
\begin{equation}\label{pilot_matrix_24}
\tilde{\mathbf{P}}_{k,g}=\begin{bmatrix}
\mathbf{p}_{k,1}^T,\mathbf{p}_{k,2}^T\cdots,\mathbf{p}_{k,g-1}^T,\mathbf{p}_{k,g+1}^T,\cdots,\mathbf{p}_{k,G_{\text{s}}}^T\\
\mathbf{p}_{k,1}^H,\mathbf{p}_{k,2}^H\cdots,\mathbf{p}_{k,g-1}^H,\mathbf{p}_{k,g+1}^H,\cdots,\mathbf{p}_{k,G_{\text{s}}}^H
\end{bmatrix}^T.
\end{equation}
By replacing $\hat{\mathbf{h}}_{k-1}$ with $\hat{\mathbf{h}}^{(n)}_{k,g}$ in (\ref{llr3}), the LLR values of the $g$-th subblock are re-calculated, denoted as $\eta_{k,g}(i,\hat{\mathbf{h}}^{(n)}_{k,g})$  for $i=1,2,\cdots,l$. Then (\ref{index_cal}) is applied to these latest LLR values, where the pilot indices of the $g$-th subblock are updated as $\hat{\mathcal{I}}^{(n)}_{k,g}=\{\hat{i}_{k,g}^{(n)}(1),\hat{i}_{k,g}^{(n)}(2),\cdots,\hat{i}_{k,g}^{(n)}(l_\text{p})\}$, satisfying $\hat{i}_{k,g}^{(n)}(1)<\hat{i}_{k,g}^{(n)}(2)<\cdots<\hat{i}_{k,g}^{(n)}(l_\text{p})$. By repeating the operations above on each of the $G_{\text{s}}$ subblocks, the output index pattern of the $k$-th data block in the $n$-th iteration can be obtained, expressed as $\hat{\mathcal{I}}_k^{(n)}=\{\hat{\mathcal{I}}^{(n)}_{k,1},\hat{\mathcal{I}}^{(n)}_{k,2},\cdots,\hat{\mathcal{I}}^{(n)}_{k,G_{\text{s}}}\}$.

With the iterations going on, when $\hat{\mathcal{I}}_k^{(n)}$ equals $\hat{\mathcal{I}}_k^{(n-1)}$, it is implied that the detection of the pilot index pattern has converged. Therefore, the stopping criterion for the iterations is set as $\hat{\mathcal{I}}_k^{(n)}=\hat{\mathcal{I}}_k^{(n-1)}$. By defining $\hat{\mathcal{I}}_k$ as the final output of the proposed turbo receiving algortihm for the $k$-th data block, the fine channel estimate, denoted as $\hat{\mathbf{h}}_k$, can be calculated using LS criterion as (\ref{eq_23_ls}) with all the detected flexible pilots corresponding to $\hat{\mathcal{I}}_k$. Afterwards, $\hat{\mathcal{I}}_k$ and $\hat{\mathbf{h}}_k$ for $k=1,2,\cdots,G$ are fed into the subsequent stages for signal demodulation. Besides, $\hat{\mathbf{h}}_k$ is also employed as the prior knowledge of CSI for coarse index pattern detection of the $(k+1)$-th data block. The aforementioned turbo receiving algorithm can be summarized as Algorithm \ref{algo1}, where the initialization phase is omitted for brevity.

{\bf Remark 1:} As illustrated in (\ref{eq22})-(\ref{pilot_matrix_24}), $\hat{\mathcal{I}}^{(n-1)}_{k}\setminus \hat{\mathcal{I}}^{(n-1)}_{k,g}$ is employed as the extrinsic information in the $n$-th iteration to update the index pattern of the $g$-th subblock, instead of $\hat{\mathcal{I}}^{(n-1)}_{k}$. Such arrangement is to eliminate the positive feedback that hinders the convergence of the proposed receiving algorithm, inspired by the principles of iterative decoding for turbo codes.

{\bf Remark 2:} For the $k$-th and $(k+1)$-th data blocks, when the physical channel coefficients and I/Q imbalance parameters remain approximately unchanged, it is reasonable to employ $\hat{\mathbf{h}}_k$ as the prior CSI knowledge of the $(k+1)$-th data block for coarse index pattern detection, since there only exists a relatively small phase noise difference between $\mathbf{h}_k$ and $\mathbf{h}_{k+1}$. On the other hand, for fast-fading channel, where the physical channel coefficient also varies across adjacent data blocks, $\hat{\mathbf{h}}_{k}$ is considered to be outdated for the $(k+1)$-th block. Under such ``worst'' case, it is validated that the proposed turbo receiving algorithm is still capable of achieving superior performance of index pattern detection and channel estimation. Aside from the relevant simulation results, an intuitive geometrical analysis is provided below to support the validity of the proposed receiving algorithm under relatively fast-fading channel.

\begin{algorithm}[t!]
\caption{Proposed Turbo Receiving Algorithm for the THz Communication System with flexible IM-aided Pilots}
\label{algo1}
\begin{algorithmic}[1]
\Require
 Received signals $y_{k,g}(i)$, initial channel estimate $\hat{\mathbf{h}}_0$, flexible pilots $\mathbf{p}_{k,g}$, flexible pilot matrixes $\tilde{\mathbf{P}}_{k,g}$ defined as (\ref{pilot_matrix_24}), distortion plus noise power $(\kappa^2\hat{P}_\text{r}+\sigma^2)$, constellation alphabets $\mathcal{M}_\text{s}$ and $\mathcal{M}_{\text{p}}$ and their sizes $M_\text{s}$ and $M_\text{p}$, block number $G$, subblock size $l$, subblock number $G_{\text{s}}$;
\Ensure
 $\hat{\mathcal{I}}_k$ and $\hat{\mathbf{h}}_k$ for $k=1,2,\cdots,G$;
\For {($k=1;k\leq G;k++)$}
\For {($g=1;g\leq G_\text{s};g++)$}
\State Compute $\eta_{k,g}(i,\hat{\mathbf{h}}_{k-1})$ using (\ref{llr3}) with $\hat{\mathbf{h}}_{k-1}$ for \Statex \quad \quad\quad $i=1,2,\cdots,l$;
\State $\hat{\mathcal{I}}_{k,g}^{(0)}=\arg \underset{{1\leq i\leq l}} {l_\text{p}\text{-}\max}\:\eta_{k,g}(i,\hat{\mathbf{h}}_{k-1})$;
\EndFor
\State $\hat{\mathcal{I}}^{(0)}_{k}=\{\hat{\mathcal{I}}_{k,1}^{(0)},\hat{\mathcal{I}}_{k,2}^{(0)},\cdots,\hat{\mathcal{I}}_{k,G_{\text{s}}}^{(0)}\}$;
\State $n=0$;
\Repeat
\State $n++$
\For {($g=1;g\leq G_\text{s};g++)$}
\State Update $\tilde{\mathbf{y}}_{k,g}$ using (\ref{eq22}) with $\hat{\mathcal{I}}^{(n-1)}_{k}\setminus \hat{\mathcal{I}}^{(n-1)}_{k,g}$;
\State $\hat{\mathbf{h}}^{(n)}_{k,g}=(\tilde{\mathbf{P}}_{k,g}^H\tilde{\mathbf{P}}_{k,g})^{-1}\tilde{\mathbf{P}}_{k,g}^H\tilde{\mathbf{y}}_{k,g}$;
\State Compute $\eta_{k,g}(i,\hat{\mathbf{h}}^{(n)}_{k,g})$ using (\ref{llr3}) with $\hat{\mathbf{h}}^{(n)}_{k,g}$ for
\Statex \quad \quad\quad\quad\;\;$i=1,2,\cdots,l$;
\State $\hat{\mathcal{I}}_{k,g}^{(n)}=\arg \underset{{1\leq i\leq l}} {l_\text{p}\text{-}\max}\:\eta_{k,g}(i,\hat{\mathbf{h}}^{(n)}_{k,g})$;
\EndFor
\State $\hat{\mathcal{I}}^{(n)}_{k}=\{\hat{\mathcal{I}}_{k,1}^{(n)},\hat{\mathcal{I}}_{k,2}^{(n)},\cdots,\hat{\mathcal{I}}_{k,G_{\text{s}}}^{(n)}\}$;
\Until{($\hat{\mathcal{I}}_k^{(n-1)}=\hat{\mathcal{I}}_k^{(n)}$)}
\State $\hat{\mathcal{I}}_k=\hat{\mathcal{I}}_k^{(n)}$;
\State Compute $\hat{\mathbf{h}}_k$ using (\ref{eq_23_ls}) with all the detected flexible
\Statex \quad\: pilots corresponding to $\hat{\mathcal{I}}_k$;
\EndFor
\State \Return $\hat{\mathcal{I}}_k$, $\hat{\mathbf{h}}_k$ for $k=1,2,\cdots,G$;
\end{algorithmic}
\end{algorithm}

\begin{figure*}[t!]
\centering
\subfigure[Frame structure with classical fixed preamble.]{
\begin{minipage}[b]{0.6\textwidth}
\includegraphics[width=1\textwidth]{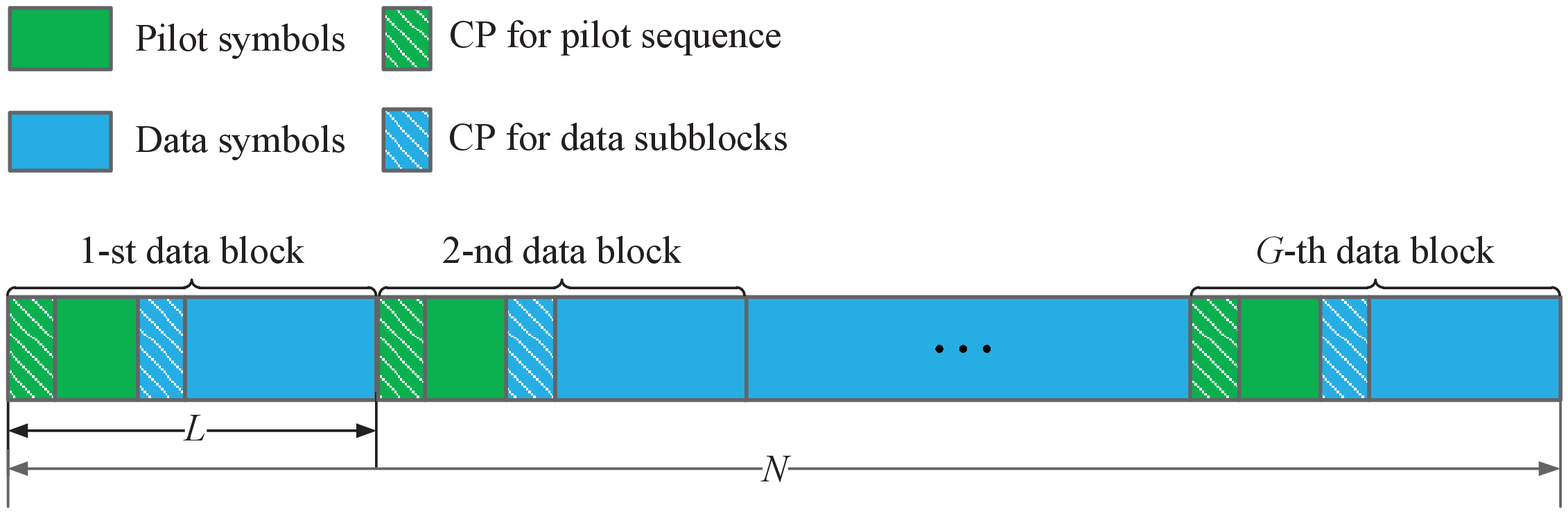} \\
\end{minipage}
}
\subfigure[Frame structure with the proposed flexible IM-aided pilots.]{
\begin{minipage}[b]{0.6\textwidth}
\includegraphics[width=1\textwidth]{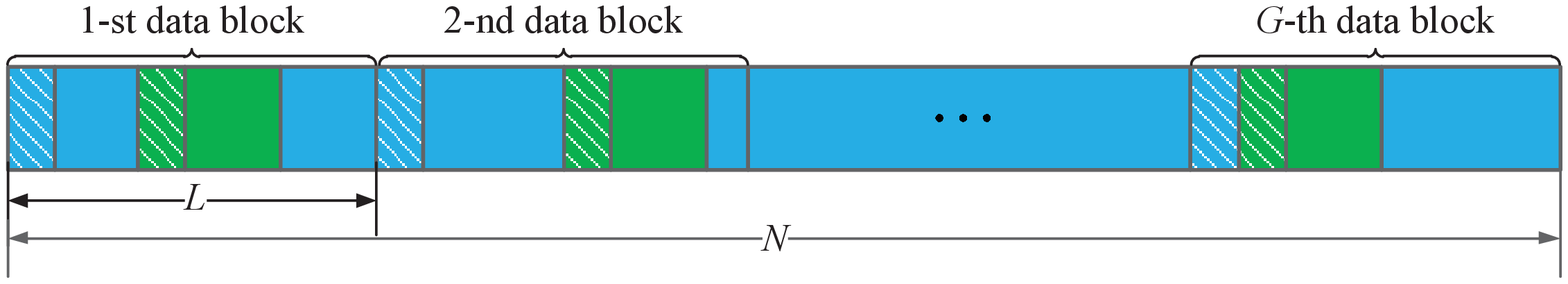} \\
\end{minipage}
}
\caption{Frame structure with classical fixed preamble and the proposed flexible IM-aided pilots under frequency-selective fading channel.}
\label{figr1}
\end{figure*}

{
\subsection{Extension to Frequency-Selective Fading Channel}
In this paper, flat-fading channel is mainly considered for THz communication systems with high-gain directional antennas \cite{Han_tsp_16}. However, the multi-path effects in the THz band, which are explicitly described in \cite{Han_twc_multipath_15}, cannot be fully eliminated by highly directional transmission under certain cases \cite{Peng_twc_19}. For instance, there may exist possible reflectors close to the LoS path, yielding additional reflected paths; Besides, combined reflection and transmission of THz signals on a multi-layer objects like window glass also contribute to multiple available paths. Furthermore, when the communication bandwidth is sufficiently high, frequency-selective channel fading should be considered even for LoS scenario without multi-path propagation \cite{Schram_conf_20}. Hence, in this subsection, the proposed flexible IM-aided pilot design is extended to the frequency-selective fading channel, where some guidelines about the generalized design will be provided.

Several assumptions are made under the frequency-selective fading channel: a) The length of channel impulse response (CIR) is assumed to be known, denoted as $L_h$; b) The CIR keeps approximately constant within each data block of length $L$, since the symbol period is extremely short due to the ultra-high communication bandwidth. Under these assumptions, the generalized data block structure with the proposed flexible IM-aided pilots is modified as illustrated in Fig. \ref{figr1}. On one hand, for each data block of length $L$ with classical fixed preamble in Fig. \ref{figr1} (a), a pilot sequence is added at the beginning of the block, consisting of the cyclic prefix (CP) part of length $L_c\geq L_h$ and pilot symbol part with length $L_p$. By considering the existence of I/Q imbalance at the transmitter, $L_p$ should be no less than $2L_h$ for channel estimation \cite{Jiang_cl_13}. Besides, CP of length $L_c$ is also applied to the data symbol part. On the other hand, unlike its classical counterpart, the pilot sequence is no longer fixed in each of the data block with the proposed flexible IM-aided pilots, as shown in Fig. \ref{figr1} (b). By such operations, additional information bits can be conveyed by the positions of pilots to compensate for the spectral efficiency loss caused by channel estimation, which have totally $(L-2L_c-L_p+1)$ possible choices for each data block. Then each data block could convey additional $\left \lfloor \log_2(L-2L_c-L_p+1) \right \rfloor$ index bits, yielding the system spectral efficiency calculated as
\begin{equation}
\begin{aligned}
{\text{SE}}'_{\text{prop}}=&\frac{1}{L}\big ((L-2L_c-l_\text{p})\log_2(M_{\text{s}})+\\&\left \lfloor \log_2(L-2L_c-L_p+1) \right \rfloor \big ),
\end{aligned}
\end{equation}
where $M_{\text{s}}$ denotes the modulation order of data symbols. At the receiver, for the $k$-th data block, the index pattern can be coarsely estimated using the channel estimate of the $(k-1)$-th data block as the prior knowledge of CSI. Note that a fixed pilot sequence is required to provide initial channel estimate for the $1$-st data block, as stated in the Initialization phase in Section \ref{s3.3}. Before coarse index pattern detection, equalization techniques such as frequency-domain equalization (FDE) or decision-feedback equalization (DFE) based on zero-forcing (ZF) or minimum mean squared error (MMSE) criterions are employed to eliminate the inter-symbol interference (ISI) caused by the frequency selectivity of THz channels \cite{Schram_conf_20}. Afterwards, a correlation-based method could be utilized to detect the index of pilot sequences. More specifically, sliding correlation is performed on the equalized signal vector, denoted as $\hat{\mathbf{x}}_k=\left[\hat{x}(1),\hat{x}(2),\cdots,\hat{x}(L-L_c)\right]$, using the pilot sequence (including CP) ${\mathbf{p}}'_k=\left[{p}'_k(1),{p}'_k(2),\cdots,{p}'_k(L_c+L_p)\right]$, which yields cross-correlation values $R_k[n]$ for $n=1,2,\cdots,L-2L_c-L_p+1$ as
\begin{equation}
R_k[n]=\sum_{k=1}^{L_c+L_p}{p}'(k)\hat{x}_k^*(n+k-1),\:\:n=1,2,\cdots,L-2L_c-L_p+1.
\end{equation}
Then the start index of the pilot sequence in the $k$-th data block can be obtained as
\begin{equation}
\hat{n}_k=\arg\underset{1\leq n\leq L-2L_c-L_p+1}{\max} \left \| R_k[n] \right \|^2.
\end{equation}
Finally, the index bits can be recovered from $\hat{n}_k$ using a look-up table between different index patterns and binary bits, and the demodulation of data symbols is straightforward. The aforementioned extension of the proposed flexible IM-aided pilot design to the THz frequency-selective fading channel will be discussed into more detail in our near-future research.}
\section{Performance Analysis}\label{s4}
\subsection{Geometrical Analysis}
\begin{figure}[t!]
\begin{center}
\includegraphics[width=0.6\linewidth, keepaspectratio]{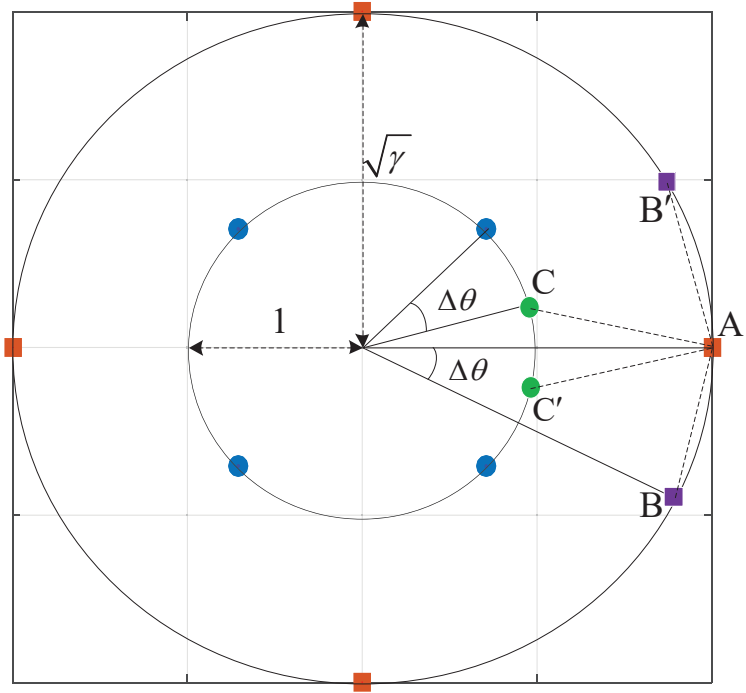}
\end{center}
\caption{Constellation diagram for the geometrical analysis, where the rotation angle corresponding to ${\text{B}}'$ and ${\text{C}}'$ under the case of $\Delta\theta\in(\pi/4,\pi/2)$ is not marked.}
\vspace{-5mm}
\label{fig5}
\end{figure}

\begin{figure}[t!]
\begin{center}
\includegraphics[width=0.7\linewidth,keepaspectratio]{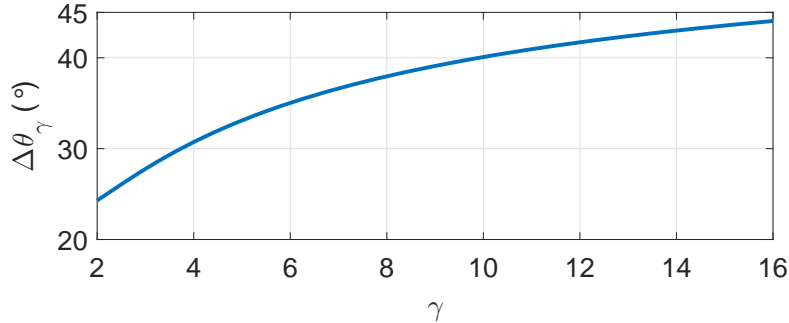}
\end{center}
\caption{Solutions to (\ref{eq_deltatheta}) with respect to the value of $\gamma$.}
\label{fig6}
\end{figure}

Under relatively fast-fading channel, there exists apparent difference between the physical channel coefficients of the $k$-th and $(k+1)$-th data blocks for $k=1,2,\cdots,G-1$, aside from the variations of phase noise. Then $\hat{\mathbf{h}}_k$, which is served as the prior CSI knowledge for the $(k+1)$-th data block, is likely to be outdated. To validate the feasibility of the proposed turbo receiving algorithm, an intuitive geometrical demonstration is provided to figure out the error probability of the LLR-based index pattern detection. For tractable analysis, several assumptions are made as below: (1) $v_\text{t}$ is approximated as zero; (2) $\mathcal{M}_\text{s}=\left \{e^{j\frac{1}{4}\pi},e^{j\frac{3}{4}\pi},e^{j\frac{5}{4}\pi},e^{j\frac{7}{4}\pi}\right \}$ and $\mathcal{M}_\text{p}=\left \{ \sqrt{\gamma},\sqrt{\gamma}j,-\sqrt{\gamma},-\sqrt{\gamma}j \right \}$ are employed for modulation, where $\gamma$ is empirically set to be larger than $2$; (3) There is only phase difference in the physical channel coefficients between adjacent blocks, whilst the amplitude remains approximately unchanged\footnote{Due to the ultra-wide bandwidth employed in the THz communication system, the change of transmission distance is marginal across hundreds of symbol periods, where the path gain is approximately constant.}. Then the received signal model in (\ref{eq8}) can be simplified as (the indices are omitted for brevity)
\begin{equation}\label{eq25}
y\approx he^{j\theta_\text{t}}\mu_\text{t}x+\tilde{w}=\tilde{h}x+\tilde{w}.
\end{equation}
Without loss of generality, we consider a pilot symbol $x=S_\text{p}(1)$ is transmitted, and set the corresponding channel fading coefficient as $\tilde{h}=1$. Then the outdated prior knowledge of CSI is written as ${\tilde{h}}'=\exp(-j\Delta \theta)$, where $\Delta \theta$ is phase difference from the actual channel, and can be set within $(0,\pi/2)$ due to the rotational symmetry of the PSK constellation alphabets. Based on this model, the LLR value for the pilot symbol $x$, denoted as $\eta$, is calculated as
\begin{equation}\label{llr4}
\begin{aligned}
\eta\approx&\log \left(\frac{l_\text{p}M_{\text{s}}}{M_\text{p}(l-l_\text{p})}\right)+\log\left (\sum_{m=1}^{M_\text{p}}\exp\left ( -\frac{\left \|y- S_\text{p}(m)e^{-j\Delta\theta} \right \|^2}{\kappa^2\hat{P}_\text{r}+\sigma^2} \right )  \right )-\\&\log\left (\sum_{n=1}^{M_\text{s}}\exp\left ( -\frac{\left \|y- S(n)e^{-j\Delta\theta} \right \|^2}{\kappa^2\hat{P}_\text{r}+\sigma^2} \right )  \right ).
\end{aligned}
\end{equation}
When the signal-to-distortion-and-noise ratio (SDNR) corresponding to the thermal noise and hardware imperfections at the receiver is sufficiently high, (\ref{llr4}) can be further approximated as
\begin{equation}\label{llr5}
\begin{aligned}
\eta&\approx\log\left ( \frac{l_\text{p}M_{\text{s}}}{M_\text{p}(l-l_\text{p})}\right)+\log\left (\underset{1\leq m\leq M_{\text{p}}}{\max}\exp\left ( -\frac{\left \|y- S_\text{p}(m)e^{-j\Delta\theta} \right \|^2}{\kappa^2\hat{P}_\text{r}+\sigma^2} \right )  \right )\\&-\log\left (\underset{1\leq n\leq M_{\text{s}}}{\max}\exp\left ( -\frac{\left \|y- S(n)e^{-j\Delta\theta} \right \|^2}{\kappa^2\hat{P}_\text{r}+\sigma^2} \right )  \right ).
\end{aligned}
\end{equation}
It is seen from (\ref{llr5}) that, at high-SDNR region, the accuracy of index pattern detection is mainly dependent on the minimum Euclidean distances between $y$ and the constellation points in $\mathcal{M}_\text{p}$ and $\mathcal{M}_\text{s}$ with a clockwise rotation of $\Delta\theta$. For the sake of clarity, the employed constellation alphabets are drawn in Fig. \ref{fig5}, where the point A denotes the received signal $y$ for sufficiently high SDNRs. Then it can be observed that
\begin{equation}\label{eq_pa_25}
\begin{aligned}
\arg\underset{S_\text{p}(m)\in\mathcal{M}_\text{p}}{\min}\left \|y- S_\text{p}(m)e^{j\Delta\theta} \right \|^2= \left\{\begin{matrix}
S_\text{p}(1), &\Delta\theta\in(0,\pi/4];  \\
S_\text{p}(2), &\Delta\theta\in(\pi/4,\pi/2),
\end{matrix}\right.
\end{aligned}
\end{equation}
and
\begin{equation}\label{eq_pa_26}
\arg\underset{S(n)\in\mathcal{M}_\text{s}}{\min}\left \|y- S(n)e^{j\Delta\theta} \right \|^2= S(1).
\end{equation}
According to (\ref{eq_pa_25}) and (\ref{eq_pa_26}), when $\Delta\theta\in(0,\pi/4]$, the pilot symbol $x$ is likely to be wrongly detected if point A is closer to C than B, where B and C correspond to $S_\text{p}(1)e^{-j\Delta\theta}$ and $S(1)e^{-j\Delta\theta}$, respectively. Define $\Delta\theta_\gamma$ as the value of $\Delta\theta$ within $(0,\pi/4]$ satisfying $|\text{AB}|=|\text{AC}|$, then it can be seen from Fig. \ref{fig5} that the sub-interval of $\Delta\theta$ in $(0,\pi/4]$ meeting $|\text{AB}|>|\text{AC}|$ yields $[\Delta\theta_\gamma,\pi/4]$. Besides, $\Delta\theta_\gamma$ can be obtained by solving
\begin{equation}\label{eq_deltatheta}
-2\gamma\cos\Delta\theta_\gamma+2\sqrt{\gamma}\cos(\pi/4-\Delta\theta_\gamma)+\gamma-1=0,
\end{equation}
whose solution within $(0,\pi/4]$ with respect to the value of $\gamma$ is presented in Fig. \ref{fig6}.

On the other hand, for $\Delta\theta\in(\pi/4,\pi/2)$, ${\text{B}}'$ and ${\text{C}}'$ correspond to $S_\text{p}(2)$ and $S(1)$ with a clockwise phase rotation of $\Delta\theta$, respectively. Similar to $\Delta\theta\in(0,\pi/4]$, $x$ is likely to be wrongly detected if A is closer to ${\text{C}}'$ than ${\text{B}}'$. Due to the symmetric property of the constellation alphabets, the sub-interval of $\Delta\theta$ that meets $|\text{A}{\text{B}}'|>|\text{A}{\text{C}}'|$ can be readily obtained as $(\pi/4,\pi/2-\Delta\theta_\gamma)$. Therefore, the overall interval width of $\Delta\theta$ causing wrong detection of $x$ at high-SDNR region, termed as \emph{wrong detection region}, can be approximated as $\pi/2-2\Delta\theta_\gamma$. For instance, when $\gamma\geq 4$, it is seen from Fig. \ref{fig6} that the width of wrong detection area is less than $\pi/6$, which means the possibility for correct index pattern detection approximately exceeds $2/3$, if $\Delta\theta$ is uniformly distributed in $(0,\pi/2)$. Hence, it can be concluded that, when the SDNR is sufficiently high, with proper design of $\mathcal{M}_\text{s}$ and $\mathcal{M}_\text{p}$, the indices of flexible IM-aided pilots can be correctly detected even using outdated CSI knowledge, which enables the proposed turbo receiver to gradually enhance the precision of the index pattern detection through iterations.
\subsection{Complexity Analysis}
For classical pilot assignment, according to (\ref{LS}), the computational complexity of channel estimation can be calculated as $2L_{\text{pre}}$ in terms of the number of complex multiplications per data block. On the other hand, for the proposed flexible IM-aided pilot design, the complexity of the proposed turbo receiving algorithm is investigated. Since the LS channel estimation in the initialization phase only performs once before transmission, it can be omitted in the calculation of detection complexity. Besides, for coarse index pattern detection, $3(M_\text{p}+M_\text{s})L$ complex multiplications are required for each data block. Moreover, for iterative joint index pattern detection and channel estimation, in each iteration, the numbers of complex multiplications for the LLR detector and LS channel estimator per data block are equal to $3G_\text{s}(M_\text{p}+M_\text{s})l$ and $2G_\text{s}(L_\text{p}-l_\text{p})$, respectively. Hence, the total count of required complex multiplications per data block yields $3(1+n_{\text{iter}})(M_\text{p}+M_\text{s})L+2n_{\text{iter}}(G_\text{s}-1)L_\text{p}$, where $n_{\text{iter}}$ represents the iteration times. Although additional overhead is induced by the proposed turbo receiving algorithm compared with the conventional counterpart, $n_{\text{iter}}$ is usually less than $4$ for fast fading channel according to the numerical results in Section \ref{s5}, which is expected to be further reduced under the scenario of quasi-static fading channel. Besides, the number of complex multiplications per data block only increases linearly with the system parameters including $M_\text{p}$, $M_\text{s}$, $G_\text{s}$, $L$ and $L_\text{p}$. Therefore, the complexity of the proposed turbo receiving algorithm is acceptable, considering its remarkable benefits over classical pilot assignment.
\begin{figure}[t!]
\begin{center}
\includegraphics[width=0.8\linewidth, keepaspectratio]{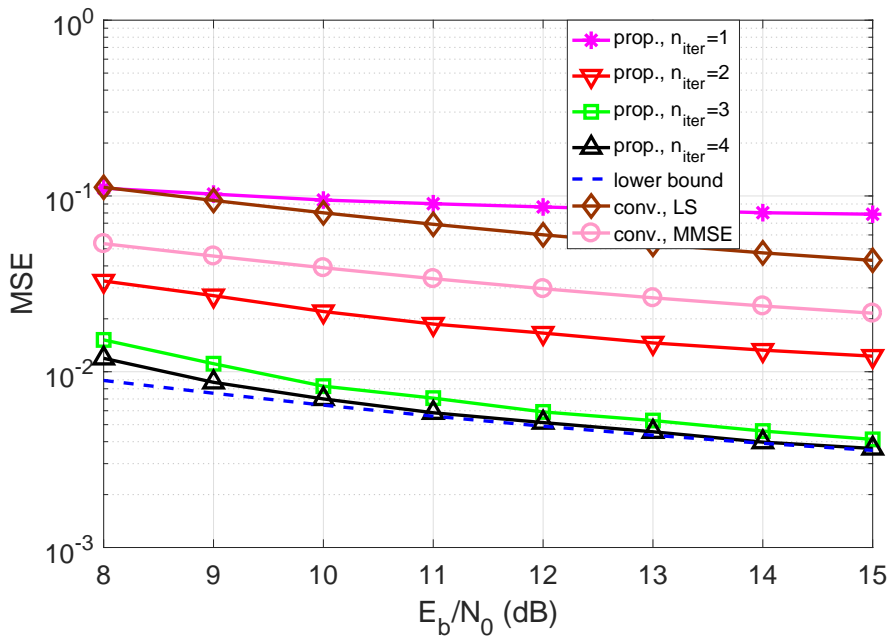}
\end{center}
\caption{Channel estimation accuracy comparison between the proposed flexible IM-aided pilot design and classical fixed pilot assignment with different number of iterations for the turbo receiving algorithm. }
\label{fig7}
\end{figure}
\section{Numerical Results}\label{s5}
The channel estimation accuracy and BER performance of the THz communication system with the proposed flexible IM-aided pilots and turbo receiving algorithm are evaluated via Monte Carlo simulations, compared with its conventional counterpart with fixed pilot assignment. Besides, the latency induced by the proposed turbo receiver is also investigated. The parameters used in simulations are listed as below. {A single-carrier THz communication system at 300 GHz with the bandwidth of 10 GHz is considered, where the transmission distance is set as $5$ m \cite{Han_tsp_16}.} In simulations, the amplitude of the physical channel fading coefficient is assumed to be approximately constant, whilst the phase term randomly changes across different data blocks, which keeps constant within each block, following independent uniform distribution between $[0,2\pi]$. For the frame structure, the block length $L$ is set as $64$, which can be further split into $G_\text{s}=8$ subblocks of length $l=8$. The preamble length of classical pilot assignment $L_\text{pre}$ equals $2$ for each data block, whilst ${L}'_{\text{pre}}=2$, $L_{\text{p}}=8$ and $l_{\text{p}}=1$ are set for the proposed pilot design. In addition, for hardware imperfections, the amplitude and phase imbalances of the I/Q branch at the transmitter are set as
$\epsilon_{\text{t}}=0.2$ and $\phi_{\text{t}}=2^{\circ}$. Besides, $\Delta\theta_{\text{t},k}$ for $k=1,2,\cdots,G-1$ is assumed to be independently identically distributed Gaussian variables following $\mathcal{N}(0,(5^{\circ})^2)$. Besides, the level of hardware impairments at the receiver is set as $-16$ dB. Note that these parameters of hardware impairments are on the same order as those in the existing literature \cite{Ramadan_access_18,Boulogeorgos_access_19,Colavolpe_jsac_05}. Moreover, in simulations, $\mathcal{M}_\text{s}=\left \{e^{j\frac{1}{4}\pi},e^{j\frac{3}{4}\pi},e^{j\frac{5}{4}\pi},e^{j\frac{7}{4}\pi}\right \}$ and $\mathcal{M}_\text{p}=\left \{ \sqrt{\gamma},\sqrt{\gamma}j,-\sqrt{\gamma},-\sqrt{\gamma}j \right \}$ are adopted by the proposed scheme, whilst QPSK is applied for both data symbols and pilots in its conventional counterpart. Based on the aforementioned parameters, the system spectral efficiencies of the proposed pilot design and its conventional counterpart are calculated as $2.125$ and $1.9375$ bit/s/Hz, respectively.

\begin{figure}[t!]
\begin{center}
\includegraphics[width=0.8\linewidth, keepaspectratio]{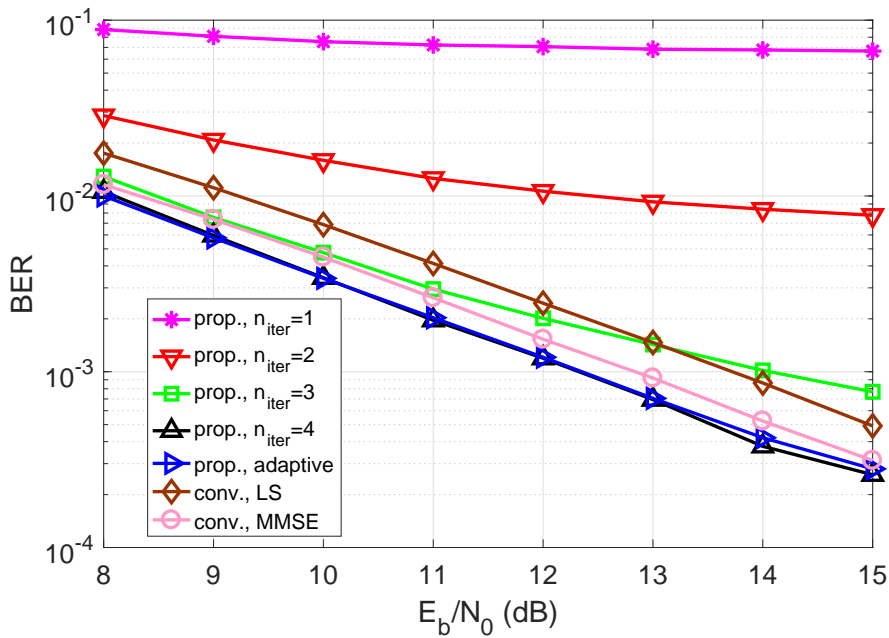}
\end{center}
\caption{BER performance comparison between the proposed flexible IM-aided pilot design with the turbo receiving algorithm and the classical fixed pilot assignment.}
\label{fig8}
\end{figure}
Figure \ref{fig7} presents the mean-square error (MSE) results for channel estimation of the proposed turbo receiving algorithm with flexible IM-aided pilots, {using LS and minimum mean-square error (MMSE) estimators with classical pilot assignment as the benchmarks.} Here $\gamma$ is set as $4$, and $E_\text{b}/N_0$ denotes the signal-to-noise ratio (SNR) per bit. A lower bound for the achievable MSE of the proposed turbo receiving algorithm is obtained by performing LS estimation with all the flexible pilots, assuming that their indices are perfectly detected by the receiver. Besides, the maximum iteration number is fixed as $n_{\text{iter}}=\{1,2,3,4\}$, respectively, where the stopping criterion for iterations stated in Section \ref{s3.3} is not employed. It is clear that the estimation MSE for the proposed turbo receiving algorithm declines with the increase of $n_{\text{iter}}$, which gradually converges to the lower bound. Explicitly, when $n_{\text{iter}}$ equals $4$, the MSE curve fits the lower bound quite well as $E_\text{b}/N_0$ is larger than $10$ dB. This demonstrates that, with the aid of the proposed turbo receiver, the flexible IM-aided pilots are capable of approaching the estimation accuracy of fixed preamble assignment with the same pilot length, whilst fully compensating for the spectral efficiency loss at the same time. Moreover, the proposed scheme improves the estimation accuracy significantly at even higher spectral efficiency over its conventional counterparts {based on LS and MMSE channel estimation}, which attains MSE of lower than $10^{-2}$ in most of the SNR region. This is because more pilot symbols can be utilized for channel estimation by employing the flexible IM-aided pilot design without sacrificing the spectral efficiency.

Besides, the BER performances of the THz communication system employing flexible IM-aided pilots and the proposed turbo receiver with different numbers of iterations are presented in Fig. \ref{fig8}, where $\gamma$ equals $4$. When $n_\text{iter}<3$, severe error floor is witnessed for the proposed scheme due to the poor detection accuracy of the pilot index pattern. With the increase of $n_\text{iter}$, the performance of the proposed scheme enhances significantly thanks to the reduced error of the index pattern detection and channel estimation, which is aligned with the MSE results in Fig. \ref{fig7}. {Besides, the proposed strategy with $n_{\text{iter}}=4$ is capable of achieving about $1.5$ dB and $0.5$ dB performance gains over its conventional counterparts using LS an MMSE channel estimators at the BER of $10^{-3}$, respectively}, whilst enhancing the spectral efficiency by about $0.2$ bit/s/Hz at the same time. This validates the remarkable advantages of the proposed flexible IM-aided pilot design and the turbo receiving algorithm in attaining desirable channel estimation accuracy and reliable high-rate data transmission. Moreover, the system performance with the proposed turbo receiver employing the iteration stopping criterion is also simulated, which achieves almost the same performance as that of $n_{\text{iter}}=4$ with reduced computational complexity and latency.

\begin{figure}[t!]
\begin{center}
\includegraphics[width=0.8\linewidth, keepaspectratio]{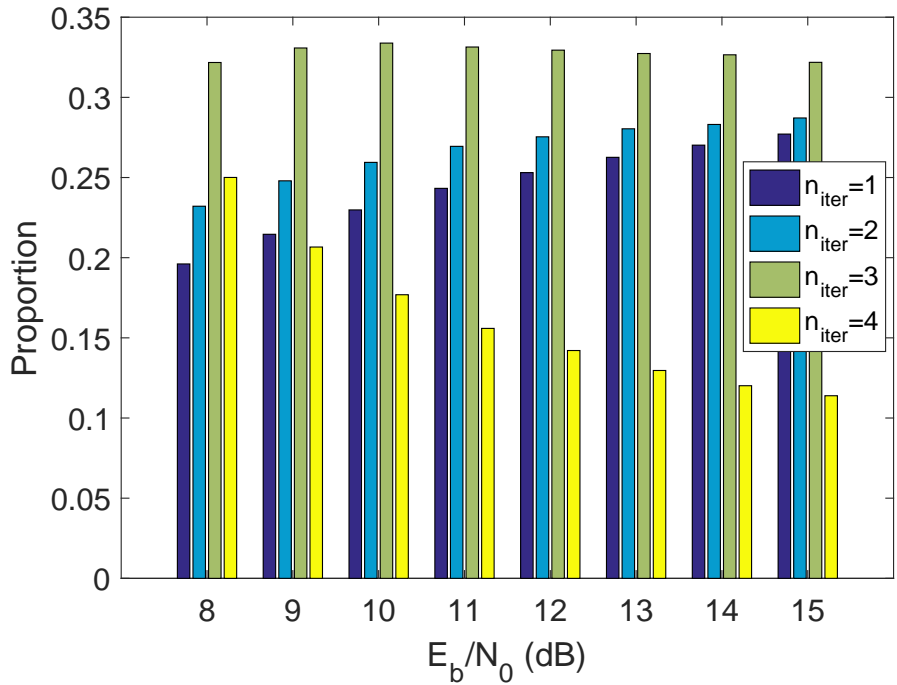}
\end{center}
\vspace{-0.5mm}
\caption{Histogram of the statistics of required iteration times for the proposed turbo receiving algorithm with the stopping criterion of iterations.}
\label{fig9}
\end{figure}

For latency analysis, the statistics of the iteration times required for the proposed turbo receiving algorithm with the stopping criterion of iterations are presented in Fig. \ref{fig9}, where the bar corresponding to $n_\text{iter}$ iterations denotes the proportion of simulated data blocks requiring $n_\text{iter}$ iterations for convergence. Besides, $\gamma$ equals $4$, and the maximal iteration number is set as $4$ for brevity. It is observed from Fig. \ref{fig9} that, although $4$ iterations are required in the turbo receiving algorithm for some data blocks, the corresponding proportion is less than $0.25$ at different SNRs. Furthermore, this proportion is gradually decreased to about $0.1$ as $E_\text{b}/N_0$ increases to $15$ dB. On the other hand, the cases of $n_\text{iter}\leq 2$ account for the proportion over $0.5$ when $E_\text{b}/N_0$ is larger than $12$ dB. Hence, it is indicated that, although additional time delay is induced by the proposed turbo receiving algorithm under fast fading channel, the resultant latency is moderate especially at high SNRs, since it is more likely to spend less than $3$ iterations for convergence. Besides, such latency is expected to be lower under the scenario of quasi-static fading channel, where the difference between the channel fading vectors of adjacent data blocks is marginal.

\begin{figure}[t!]
\begin{center}
\includegraphics[width=0.8\linewidth, keepaspectratio]{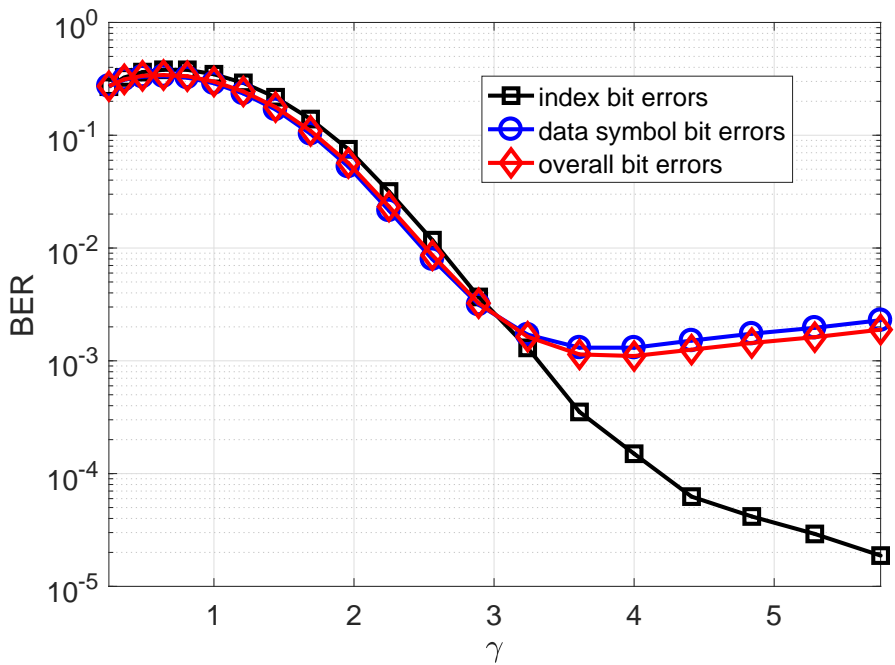}
\end{center}
\caption{BER performance of the THz communication system employing the proposed flexible IM-aided pilot design and the turbo receiving algorithm with respect to $\gamma$. }
\label{fig10}
\end{figure}

Figure \ref{fig10} shows the BER performance of the THz communication system using flexible IM-aided pilots and the turbo receiving algorithm with respect to the value of $\gamma$. It is observed that, when $\gamma\leq 1$, the overall BER value is higher than $0.2$, for the reason that the average transmit power of pilots is low, causing poor channel estimation accuracy. Furthermore, for $\gamma>1$, the BER of index bits gradually declines as $\gamma$ becomes larger, due to higher allocated power for the pilot symbols and also the enlarged distance between $\mathcal{M}_{\text{s}}$ and $\mathcal{M}_{\text{p}}$ { from $\gamma>2$ according to (\ref{dmin_gamma_cal})}. Besides, for the data symbol bits, there is also performance improvement with the increase of $\gamma$ until $\gamma\geq 4$ despite its reduced power, thanks to the enhanced accuracy of the index pattern detection and channel estimation. As $\gamma$ continues to enlarge, the performance of data part begins to deteriorate. Since the data symbol bits account for the majority of transmitted information in most cases, the trend of the overall BER with respect to $\gamma$ is generally dependent on the data symbol bits, as illustrated in Fig. \ref{fig10}. {Clearly, there is a trade-off to determine the optimal $\gamma$ for reliable data transmission, which is aligned with the analytical work in Section \ref{s3b}. Numerically, an optimal value of $\gamma$ can be approximately obtained as $4$ from the BER curve of overall bit errors.}

\section{Conclusion}\label{s6}
In THz wireless communication systems with hardware imperfections, due to the time-variant property of the strong phase noise as well as possible fluctuations of the physical channel coefficient, pilots are required to be frequently inserted for channel tracking, leading to undesirable throughput loss. To compensate for the spectral efficiency loss, whilst attaining high accuracy of channel estimation simultaneously, a novel flexible pilot design strategy is proposed based on the IM technique, where the positions of pilots are no longer fixed in the data frame, so that additional information bits could be conveyed by their indices. At the receiver, a turbo receiving algorithm is developed for iterative index pattern detection and channel estimation, which is capable of achieving superior performance even with outdated prior knowledge of CSI. To validate the feasibility of the proposed pilot design as well as the turbo receiving algorithm, an intuitive geometrical analysis is performed, followed by the computational complexity analysis. Analytical and simulative results demonstrate that the proposed flexible IM-aided pilot design with the turbo receiving algorithm could guarantee the accuracy of channel estimation. Besides, it is capable of improving the BER performance with higher spectral efficiency, in comparison with its conventional counterpart.

{Aside from the use of highly directional antennas, the multiple-input multiple-output (MIMO) architecture with beamforming/combining techniques is capable of further enhancing the coverage of the THz communication systems. For single-stream MIMO systems being quite suitable for THz wireless transmission due to low hardware cost and power consumption, the proposed flexible IM-aided pilot design is also applicable, where the beamforming vector design needs careful investigation. The detailed research will be set aside as our future work.}

\clearpage

\end{document}